\documentclass[%
preprint,
 amsmath,amssymb,
 aps,
]
{revtex4-1}
\usepackage{subfig}
\usepackage{mathrsfs}

   \usepackage[pdftex]{graphicx}
    \DeclareGraphicsExtensions{.pdf,.jpg,.png}
\usepackage{bm}
\usepackage{enumerate}
\usepackage{makecell}

\begin{document}


\title{Exploring multi-layer flow network of international trade \\ based on flow distances}

\author{Bin Shen$^{1}$}
\author{Jiang Zhang$^{2}$}
\email{zhangjiang@bnu.edu.cn} 
\author{Qiuhua Zheng$^{3}$}
\affiliation{1. Ningbo Institute of Technology, Zhejiang University, Ningbo, Zhejiang, China\\
2. School of Systems Science, Beijing Normal University, Beijing, China\\
3. School of Computer Science and Technology, Hangzhou Dianzi University, Hangzhou, China
}

\date{\today}

\begin{abstract}
Based on the approach of flow distances, the international trade flow system is studied from the perspective of multi-layer flow network. A model of multi-layer flow network is proposed for modelling and analyzing multiple types of flows in flow systems. Then, flow distances are introduced, and symmetric minimum flow distance is presented. Subsequently, we discuss the establishment of the multi-layer flow networks of international trade from two coupled viewpoints, i.e., the viewpoint of commodity flow and that of money flow. Thus, the multi-layer flow networks of international trade is explored. First, trading ``trophic levels'' are adopted to depict positions that economies occupied in the flow network. We find that the distributions of trading ``trophic levels'' have the similar clustering pattern for different types of commodity, and there are some regularities between money flow network and commodity flow network. Second, we find that active and competitive countries trade a wide spectrum of products, while inactive and underdeveloped countries trade a limited variety of products. Besides, some abnormal countries import many types of goods, which the vast majority of countries do not need to import. It may indicate an abnormal economic status. Third, harmonic node centrality is proposed and we find the phenomenon of centrality stratification. It means that competitive countries tend to occupy the central positions in the trading of a large variety of commodities, while underdeveloped countries likely in the peripheral positions in the trading of their limited varieties of products. Fourth, we find that manufactured products have significant larger mean first-passage flow distances from the source to the sink than that of primary products.
\end{abstract}

\maketitle


\section{\label{sec:intro}Introduction}
Flow network is an important tool for describing and analyzing open flow systems and has been studied extensively in a large range of flow network systems, such as ecological flow networks \cite{Gallos2007, JiangZhang_Bio2010, JiangZhang_PLoS2013}, world trade flow networks \cite{Peiteng_PLoS2014}, traffic and transportation flow networks \cite{Khan2007}.
Flow networks are commonly modelled by directed weighted networks, where directions and weights of edges represent directions and volume fluxes of flows respectively. Because open flow systems always exchange energy, matter and information with their surroundings, flow networks normally have two special nodes (i.e., the source and the sink) representing the environment, where all flows are supposed to be from the source, through the system and finally to the sink node \cite{Liangzhu2014}.

Based on the model of flow networks, many useful methods have been developed for exploring structures and dynamics of flow systems. For example, Rosvall and Bergstrom \cite{Rosvall2008} proposed a method of probability flow of random walks for revealing community structure in weighted and directed networks. V\'{e}zina and Platt \cite{Vezina1988} described an inverse method for estimating network fluxes in undersampled environments. In a recent work \cite{JiangZhang_Bio2010}, we solved a problem that how to measure the distances between nodes in flow systems. Several flow distances, i.e., the first-passage flow distance, the total flow distance and the symmetric flow distance, were put forward, and their helpfulness in calculating ``trophic levels'' and clustering for nodes has been preliminarily shown in the work.

Using these methods, some significant laws and important knowledge have already been revealed in such kinds of complex flow systems. For instance, in living things, such as animals \cite{Kleiber1932}, plants \cite{Niklas2007} and microbes \cite{Hemmingsen1950}, allometric scaling laws (e.g., the power-law relationship between an animal's metabolic rate and its body mass \cite{Kleiber1932}) exist extensively due to energy transportation on energy flow networks of living organisms \cite{West1997, Banavar1999}. The recent works have shown that the universal allometric scaling law also exists in a much broader range of flow networks, e.g., weighted food webs \cite{JiangZhang_Bio2010} and trade flow networks \cite{Peiteng_PLoS2014}.

Currently, most of studies on flow networks focus on monolayer flow networks \cite{JiangZhang_Bio2010, Peiteng_PLoS2014, JiangZhang_PLoS2013, West1997, Banavar1999}.
However, complex systems containing multi-flows cannot be well represented by monolayer flow networks. For example, in a transportation system, aircraft traffic, train traffic and highway automobile traffic may coexist;
in a city infrastructure, water flow, power flow and gas transmission flow exist at the same time; in a human body, the circulations of blood, lymph and interstitial fluid coexist. If we do not distinguish between different types of flows, and simply adopt monolayer flow network to describe these flow network systems, many implicit information of multi-flows will be lost. Thus, many important knowledge for multi-flow systems may not be revealed.
Besides flow systems having multi-flows of different types of substances, complex flow systems containing flows of the same substance with different labels (e.g., groups) also need a practical tool for system modelling and analytics. For instance, in a world wide trade system, money flows for trading different types of commodities may flow among different countries; in the logistics industry, different transportation flow networks are adopted by competing logistics enterprises. In such cases, multi-flows cannot be well distinguished with each other if we simply use monolayer flow network to model the systems. Overall, because these co-existing multi-flows are hard to be finely modelled by a monolayer flow network, it is necessary to introduce a new concept called \emph{Multi-layer Flow Networks} (MFNs). Examples of MFNs include multi-layer trade flow networks, where different types of commodities flow in different layers, and multi-layer transportation flow networks, where passenger traffic flows on the layer of air transportation network and that of rail network. To the best of our knowledge, a generalized theoretical framework of multi-layer flow networks is still lack. So it is necessary to study multi-layer flow networks and their applications in depth.

In this paper, we use world wide trade as a specific instance of multi-layer flow network. In previous work, international trade has been studied from the viewpoint of networking. Different types of trade networks (such as binary or weighted, directed or undirected ones) have been built to model real trade interactions between countries. For example, Serrano and Bogu\~{n}\'{a} \cite{Serrano2003} finds that the binary and undirected world trade web also presents the typical properties of complex networks, e.g., scale-free inhomogeneities and a high clustering coefficient. Fagiolo et al. \cite{Fagiolo2008} studied the topological properties and their evolution over time using a weighted world trade network. Fan et al. \cite{Ying2014} explored countries' roles and positions using the improved bootstrap percolation and the other methods in an international trade network of single-layer. Unlike these previous studies \cite{Serrano2003, Fagiolo2008, Ying2014, Peiteng_PLoS2014}, here we use the framework of multi-layer flow network to explore novel characteristics of international trade networks based on the methods of flow distances, which is recently proposed in \cite{Liangzhu2014}.

The rest of the paper is structured as follows. In Section 2, a formal description for multi-layer flow networks is presented, and flow distances are introduced. In Section 3, we use multi-layer flow network of international trade as a case, and discuss the establishment of the multi-layer flow network from the viewpoints of trading commodity flow and money flow respectively. Then, using the approach of flow distances, some results on countries' trophic levels, node centralities and mean first-passage flow distances from the source to the sink are discussed in Section 4. At last, we give the conclusions of this study in Section 5.

\section{Methods}  
\subsection{Modeling multi-layer flow networks}
Multi-layer flow networks (MFN) can be regarded as a special type of multi-layer networks \cite{Mikko2014, Boccaletti2014}. This tool is able to well depict multiplex flows in different layers. Thus, we have the following definition.

An multi-layer flow network is a pair \emph{MFN} = $(G, E)$, where $G = \{G_\alpha; \alpha \in {0,\ldots, M-1}\}$ is a family of directed (binary or weighted) graphs $G_\alpha = \{V, E_\alpha\}$ (called layers of \emph{MFN}), and
\begin{equation}E = \{E_{\alpha\beta} \subseteq (v, v); v \in V; \alpha,\beta\in\{0,\cdots,M-1\},
 \alpha\neq\beta\}
\end{equation}
is the set of interconnections between nodes and their counterparts in the rest of layers. The node sets $V$ in each layer are all the same, and are supposed to contain $N$ common nodes and two special nodes ``source'' and ``sink'', where ``source'' is the start of all flows and denoted as node 0, and ``sink'' is the end of all flows and labeled as node $N+1$. $E_{\alpha}$, the adjacency matrix of layer $\alpha$, can be written as follows.
\begin{equation}E_{\alpha} = \{e_{i,j}^{\alpha}\}_{(N+2)\times(N+2)}; \ \  i,j\in\{0,\cdots,N+1\},\alpha\in\{0,\cdots,M-1\}.
\end{equation}
where $e_{i,j}$ is the flow from node $i$ to node $j$. Especially, the first column and the last row of this matrix are all 0 because there are no inflow to ``source'' node and no outflow from ``sink'' node. Normally $e_{0,N+1}$ is 0, because ordinarily there is no flow from ``source'' to ``sink'' directly. The total inflow to node $j$ (denoted as $e_{\cdot j}$) is  calculated as $\sum_{i=0}^{N+1}e_{ij}$ and the total outflow from $i$ (labeled as $e_{i \cdot}$) is $\sum_{j=0}^{N+1}e_{ij}$. Because the flow system is assumed to be in equilibrium, except ``source'' and ``sink'', all other nodes have a balanced inflow and outflow, that is $e_{\cdot i}=e_{i \cdot}$, where $i=1,\cdots,N$. The flows to ``sink'' (i.e., $e_{i,N+1}$) are regarded as \emph{dissipation}.



\begin{figure}
  \centering
  \includegraphics[width=3.1in]{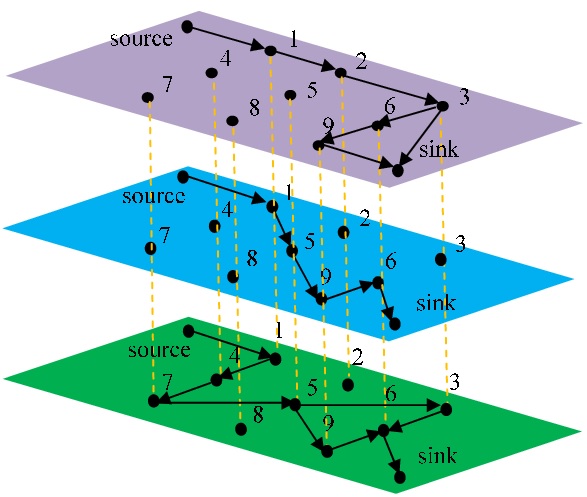}\\
  \caption{Illustration of a multi-layer flow network}
\end{figure}

%

\subsection{Distance on multi-layer flow networks}\label{Distance_subsection}
In \cite{Liangzhu2014}, L. Guo et al. have proposed flow distances on open flow networks. In this subsection, we will extend them to those on multi-layer flow networks.

Let us consider a single layer $\alpha$ of a given \emph{MFN} firstly. Suppose a large enough number (say $\lambda$) of particles flow along directed edges randomly. This random walk process can be assumed as a Markov chain. \emph{The mean first-passage flow distance} (MFPFD) from one node $i$ to another node $j$ (denoted as $l_{ij}$) is defined as the expected number of steps for reaching $j$ for the first time, given that initially the particles are at $i$. \emph{The mean total flow distance} (MTFD) (denoted as $t_{ij}$) is the average number of steps for arriving at $j$ regardless of whether it is the first time of arriving, also given that the particles are at $i$ initially.

To illustrate the above concepts vividly, the following thought experiment is designed and considered. Suppose all particles are initially no color. There are two situations.
1) If particles go through node $i$, their color turns blue; then when blue particles arrive at node $j$, they become red. In this situation, MFPFD from $i$ to $j$ is the average number of steps of blue particles turning red.
2) If particles go through node $i$, their color turns blue; blue particles remain blue, when arriving at node $j$. Then, MTFD from $i$ to $j$ is the average number of steps of all accumulated blue particles (also containing those whose arrival times is more than 1) arriving $j$. Here, a blue particle may be counted repeatedly if it arrives at $j$ for multiple times.

Based on the above experimental description, the computation of flow distances are given as below.
Formally, we have a stochastic matrix describing the transitions of the Markov chain $M_\alpha = \{m^\alpha_{ij}\}_{(N+2)(N+2)}$, where
\begin{equation}
m^\alpha_{ij}=\left\{ \begin{array}{lc} \frac{e_{ij}^{\alpha}}{\sum^{N+1}_{j=0}{e_{ij}^{\alpha}}} & \sum^{N+1}_{j=0}{e_{ij}^{\alpha}}\neq 0\\
0 & \sum^{N+1}_{j=0}{e_{ij}^{\alpha}}=0. \end{array} \right.
\end{equation}
Then, for this absorbing Markov chain, its fundamental matrix is defined as below.
\begin{equation}
U_\alpha = (I-M_\alpha)^{-1} = I + M_\alpha + M_\alpha^2 + \cdots
\end{equation}
where $I$ is the corresponding identity matrix with the same size of $M_\alpha$. Thus, we have the matrix of MTFD $T_\alpha=\{t^\alpha_{ij}\}_{(N+2)(N+2)}$ and that of MFPFD $L_\alpha=\{l^\alpha_{ij}\}_{(N+2)(N+2)}$, where
\begin{eqnarray}
t^\alpha_{ij} & = & \frac{(M_\alpha U_\alpha^2)_{ij}}{u_{ij}}, \\
l^\alpha_{ij} & = & t^\alpha_{ij} - t^\alpha_{jj}.
\end{eqnarray}
The detailed derivations of the above formulas can be found in \cite{Liangzhu2014}. Because $t^\alpha_{ij}$ and $l^\alpha_{ij}$ are asymmetric and cannot well satisfy the applications which need symmetric distance metrics (e.g., clustering and generating minimum spanning tree), a symmetric flow distance $c^\alpha_{ij}$ was also introduced in \cite{Liangzhu2014}, which is given as below.
\begin{equation}
c^\alpha_{ij} = \frac{2l^\alpha_{ij} l^\alpha_{ji}}{l^\alpha_{ij}+l^\alpha_{ji}}.
\end{equation}
Here, we present an alternative symmetric flow distance called \emph{symmetric minimum flow distance} (SMFD), which is  simply calculated as below.
\begin{equation}
f^\alpha_{ij} = \textrm{min}\{l^\alpha_{ij},\ l^\alpha_{ji}\}.
\end{equation}
These elements $f^\alpha_{ij}$ form the matrix of SMFD $F_\alpha=\{f^\alpha_{ij}\}_{(N+2)(N+2)}$.
Thus, matrices of all layers of $T_\alpha$, $L_\alpha$ and $F_\alpha$ can be assembled into the corresponding vectors, i.e.,
$V_{T} = \{T_0, T_1, \cdots,T_{M-1}\}$, $V_{L} = \{L_0, L_1, \cdots,L_{M-1}\}$ and $V_{F} = \{F_0, F_1, \cdots,F_{M-1}\}$.

Based on the matrices of MTFD, MFPFD and SMFD, many interesting knowledge can be found in the open flow systems, such as ``trophic levels'' and centralities of nodes.

\section{An empirical study: the multi-layer flow network of international trade}
\subsection{Dataset}
We use the NBER-United Nations trade data (http://cid.econ.ucdavis.edu/nberus.html) to explore new features of multi-layer flow network of international trade. The dataset covers the details of world trade flow from 1962 to 2000, and SITC4 (4-digit Standard International Trade Classification, Revision 4) standard \cite{SITC4} is used to organize hundreds types of products in the dataset. A fragment of the dataset is shown in Table \ref{table_dataset}, where ``ICode'' and ``ECode'' are the corresponding codes for importers and exporters respectively, and ``Value'' means the value of commodity whose unit is thousands of US dollars. ``DOT'' (direction of trade) has two options: 1 (data from the importer) and 2 (data from that exporter). The ``Quantity'' of commodity is measured by ``Unit'' whose codes can be ``W'' (weight of metric tons), ``V'' (Volume of cubic meters) and so on. Readers can refer to \cite{Feenstra2005} for a detailed description of this dataset.

In this study, we use the value of commodity as the volume flux of the trade.

\begin{table}
\caption{A fragment of the NBER-United Nations trade dataset}
\label{table_dataset}
\centering
  \begin{tabular}{cccccccccc}
  \hline\hline
  Year & ICode & Importer & ECode  & Exporter & SITC4 & Unit & DOT & Value & Quantity \\
  \hline
  2000 & 457640 & Thailand & 532760  & Germany & 6832 & W & 1 & 157 & 9 \\
  2000 & 457640 & Thailand & 532760  & Germany & 6842 & W & 1 & 11524 & 2731 \\
  ... & ... & ... & ...  & ... & ... & ... & ... & ... & ... \\
  2000 & 532080 & Denmark & 211240  & Canada & 8211 & W & 1 & 163 & 16 \\
  \hline\hline
\end{tabular}
\end{table}

\subsection{Establishment of the multi-layer flow network of international trade}
We use the data in year 2000 to build a multi-layer flow network of international trade. In this multi-layer flow network, there are totally 192 nodes (containing two special nodes ``source'' and ``sink'') and 1288 different layers, where each common node (and its counterparts in the different layers) represents a country and each layer contains a single product's trade flow. Trade flows can be depicted from two different perspectives: the viewpoint of trading commodity flow and that of money flow for trading. A detailed explanation is given below.

\subsubsection{Viewpoint of trading commodity flow}
In each layer, for the given product of this layer, if there exists a trade relationship between two countries, a directed edge can be built from the exporter to the importer, and the corresponding value of the trade is set as the volume flux of each edge. The edge from ``source'' to the country node (i.e., $j$) means the production of this commodity in the country $j$, and the edge from node $j$ to ``sink'' represents the consumption of this commodity in the country $j$. The volume flux of theses edges are the corresponding value of the production (or the consumption).
\begin{figure}
  \centering
  \includegraphics[width=3.3in]{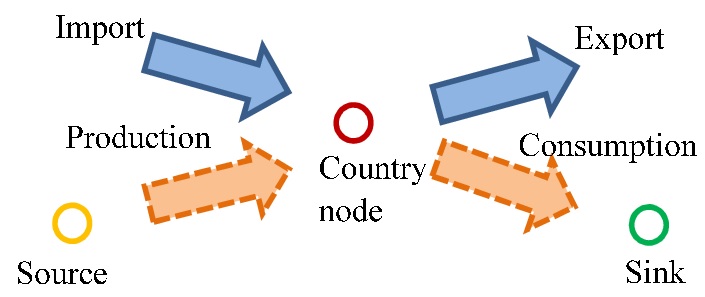}\\
  \caption{Illustration of a balanced country node from the viewpoint of commodity flow}\label{NodeBalance}
\end{figure}

For each country node $j$ (except ``source'' and ``sink'') in a certain layer $\alpha$, according to the constraint of balanced inflow and outflow (Fig. \ref{NodeBalance}), we have the following equation.
\begin{equation}
\label{balanced_flow}
\sum_{i=1}^{N}e_{ij}^{\alpha} + e_{0j}^{\alpha} = \sum_{k=1}^{N}e_{jk}^{\alpha} + e_{j,N+1}^{\alpha}
\end{equation}
where $e_{0j}^{\alpha}$ and $e_{j,N+1}^{\alpha}$ are the domestic production and consumption of country $j$ for the product in layer $\alpha$ respectively. Because $e_{0j}^{\alpha}$ (the flow from the source to $j$) and $e_{j,N+1}^{\alpha}$ (the flow from $j$ to the sink) are not available, for simplicity, they are estimated as below.
\begin{equation}
\left\{ \begin{array}{ll} e_{0j}^{\alpha}=0;\ e_{j,N+1}^{\alpha}=\sum_{i=1}^{N}e_{ij}^{\alpha}-\sum_{k=1}^{N}e_{jk}^{\alpha} & \quad \textrm{if } \sum_{i=1}^{N}e_{ij}^{\alpha}\geq\sum_{k=1}^{N}e_{jk}^{\alpha},\\
e_{j,N+1}^{\alpha}=0;\  e_{0j}^{\alpha}=\sum_{k=1}^{N}e_{jk}^{\alpha}-\sum_{i=1}^{N}e_{ij}^{\alpha} &\quad \textrm{if } \sum_{i=1}^{N}e_{ij}^{\alpha}<\sum_{k=1}^{N}e_{jk}^{\alpha}. \end{array} \right.
\label{source_sink_computation}
\end{equation}

\subsubsection{Viewpoint of money flow for trading}
Besides the viewpoint of trading commodity flow, the flow network of international trade also can be established from the viewpoint of money flow. Because the trading of commodity is always accompanied by money flow, the direction of money flow is just the opposite of that of commodity flow, that is from the importer to the exporter. 
The edge from ``source'' to the country node $j$ represents country $j$'s trade deficit driven by the consumption of the commodity, and the edge from country node $j$ to ``sink'' means the surplus of the exports over the imports.

\begin{figure}
  \centering
  \includegraphics[width=3.8in]{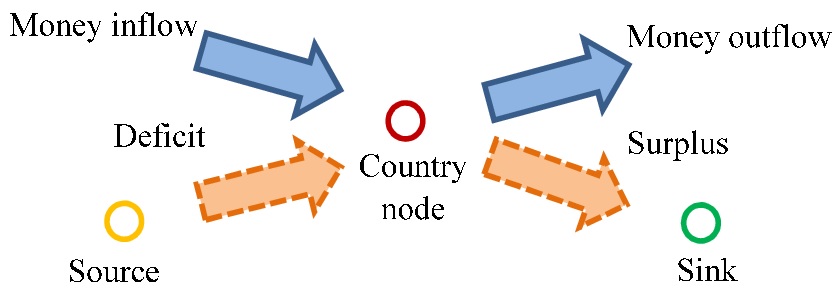}\\
  \caption{Illustration of a balanced country node from the viewpoint of money flow}\label{MoneyBalance}
\end{figure}

For each balanced country node $j$ (except the source and the sink), Equation \ref{balanced_flow} still holds (Fig. \ref{MoneyBalance}), where $e^{\alpha}_{ij}$ and $e^{\alpha}_{jk}$ are the money inflow from country $i$ to country $j$ and the money outflow from country $j$ to country $k$ respectively. $e^{\alpha}_{0j}$ and $e^{\alpha}_{j,N+1}$ are the deficit and the surplus 
respectively, which also can be calculated using Equation \ref{source_sink_computation}.

\section{Results}
\subsection{Trading trophic levels}
\subsubsection{Countries' trophic levels from the viewpoints of commodity flow and money flow}
``Trophic level'' is a term borrowed from ecology, meaning the position a species occupies in a food chain.
In flow network systems, we use this variable to indicate the position that a node occupied in the whole system, and its value is one node's distance from ``source''. Since there may be multiple paths from the source to the node, adopting the distance of the shortest path from the source may underestimate the trophic level of a node, and using the mean first-passage flow distance (MFPFD) from the source is more reasonable \cite{Liangzhu2014}.


\begin{figure}
\centering
    \subfloat[The viewpoint of commodity flow]
    {
       \begin{minipage}[t]{0.5\linewidth}
       \centering
       \includegraphics[width=3.1in]{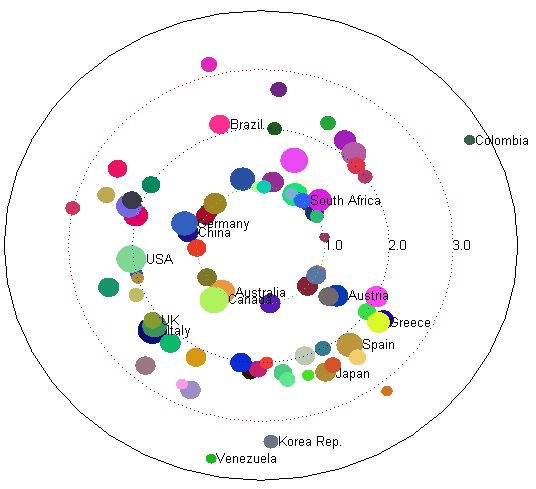}
       \vspace{0pt}
       \end{minipage}
       \label{Fig_a_viewpoint_commodity}
    }
    \subfloat[The viewpoint of money flow]{
       \begin{minipage}[t]{0.5\linewidth}
       \centering
       \includegraphics[width=2.95in]{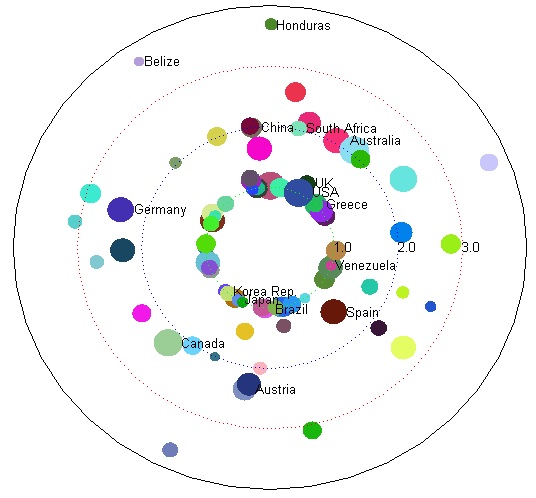}
       \vspace{0pt}
       \end{minipage}
       \label{Fig_b_viewpoint_money}
    }
    \caption{Trading trophic levels of countries for the commodity of live bovine animals from different viewpoints.}
    \label{Fig_live_bovine_animals}
\end{figure}

We use the layer of the commodity of live bovine animals as an example. Based on MFPFD discussed in the subsection \ref{Distance_subsection}, we use the MFPFD from the source to each country node as its trading trophic level. The trophic levels of each country from the viewpoints of commodity flow and money flow are given in Fig.\ref{Fig_live_bovine_animals} respectively. In the figure, country nodes are plotted in a circle, whose distances to the center of the circle are proportional to their trophic levels, and whose angles and colors are selected randomly. The nodes' sizes are in proportion to the natural logarithm of the nodes' total outflow (or inflow).

\begin{figure}
\centering
    \subfloat[The viewpoint of commodity flow]
    {
       \begin{minipage}[t]{0.5\linewidth}
       \centering
       \includegraphics[angle=0,width=3.1in]{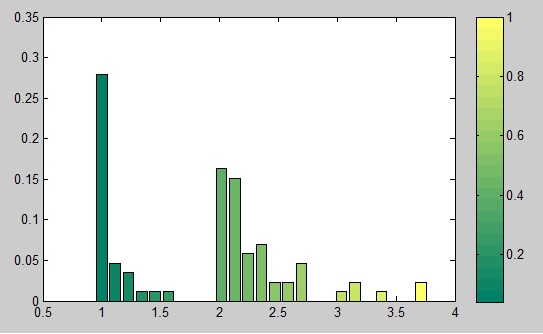}  
       \vspace{0pt}
       \end{minipage}
       \label{Fig_a_viewpoint_commodity_distribution}
    }
    \subfloat[The viewpoint of money flow]{
       \begin{minipage}[t]{0.5\linewidth}
       \centering
       \includegraphics[angle=0,width=3.1in]{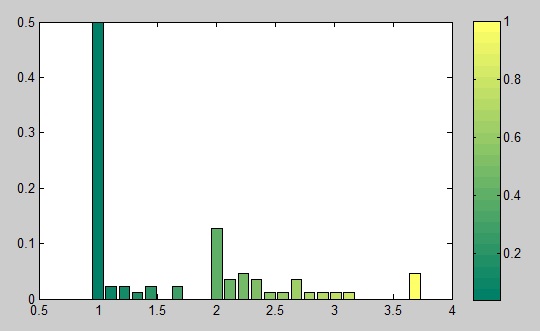}
       \vspace{0pt}
       \end{minipage}
       \label{Fig_b_viewpoint_money_distribution}
    }
    \caption{The distribution of trading trophic levels of countries for the commodity of live bovine animals}
    \label{Fig_live_bovine_animals_distribution}
\end{figure}

From the viewpoint of commodity flow, the source of commodity flow is the production of this commodity and the sink means the consumption. So for a certain commodity, the trophic level of a country node represents the position that the country occupied in the global supply chain (i.e., the commodity flow). The smaller the trophic level, the role of the country is more inclined to be the producer (i.e., the exporter) of this commodity. On the contrary, the bigger the trophic level, the greater the distance from the source to the country node and the country is more inclined to be the consumer (that is the importer) of this commodity. From Fig.\ref{Fig_a_viewpoint_commodity}, we find that some countries (such as Germany, China, Australia, Canada and South Africa), whose trading trophic levels are slightly greater than or equal to 1, are inclined to be the exporters of live bovine animals. Some other nodes, such as Korea Rep., Venezuela and Colombia, which are far away from the centre of the circle, are the importers of live bovine animals. The corresponding distribution of trophic levels of countries is given in Fig.\ref{Fig_a_viewpoint_commodity_distribution}. Obviously, there are three major groups of countries, i.e., the countries whose trophic levels are between 1 and 1.6, between 2 and 2.8, and larger than 3. The universality of this pattern has been confirmed for a large variety of commodities. 

From the viewpoint of money flow, the source is consumer demand for the commodity which leads to trade deficit, and the sink means the surplus. Because money flow can be obtained through interchanging ``source'' and ``sink'' and reversing the direction of edges of commodity flow network, it is intriguing to find some regularities and rules between money flow network (say $MG_{\alpha}$) and commodity flow network (say $CG_{\alpha}$) for the same commodity. These regularities and rules include the followings.
\begin{itemize}
\item $MG_{\alpha}$ and $CG_{\alpha}$ have the same MFPFD from the source to the sink (i.e., $l_{0,N+1}^{\alpha}$). 
    For the commodity of live bovine animals, $l_{0,N+1}^{\alpha}$ of $MG_{\alpha}$ and $CG_{\alpha}$ are both 3.3674.
\item For a country node, the larger the trophic level on $MG_{\alpha}$, the smaller on $CG_{\alpha}$, and vice versa. That is to say, given any two countries $i$ and $j$, if trophic level of $i$ is larger than that of $j$ on $MG_{\alpha}$, trophic level of $i$ is no bigger than that of $j$ on $CG_{\alpha}$, and vice versa. For instance, Venezuela has a biggest trophic level (which is 3.7134) from the viewpoint of commodity flow; on the contrary, its tropic level (which is 1.00) is the smallest from the viewpoint of money flow.
\item The distributions of trade trophic levels of countries from two viewpoints exhibit similar patterns: a large percentage of trade trophic levels are equal to or slightly larger than 1, and another group of trophic levels is between 2 and 3.
\end{itemize}

\subsubsection{Countries' trophic levels for different products}

\begin{figure}
\centering
\includegraphics[angle=0,width=5.9in]{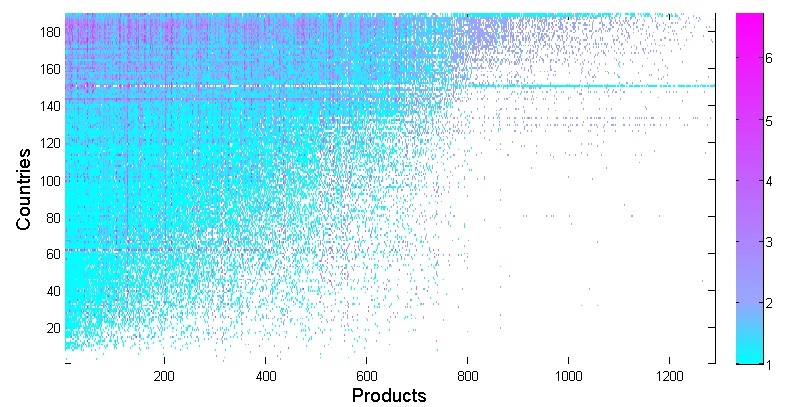}              
\caption{Countries-products matrix reporting countries' trophic levels in different layers with different products. Countries and Products are arranged in descending order of the sum of $n_{cyan}$ and $n_{magenta}$, and then $\overline{l_{0,i}^{\alpha}}$, where $n_{cyan}$, $n_{magenta}$ and $\overline{l_{0,i}^{\alpha}}$ are the number of cyan points, that of magenta points and the mean of MFPFD from the source to the country represented by the row respectively.}
\label{matrix_country-product}
\end{figure}

We compare countries' trophic levels in different layers with different products.
We portray them using the countries-products matrix as shown in Fig.\ref{matrix_country-product}. In the figure, each point represents a trophic level with the corresponding country and product, and its value is depicted by the color, where white indicates the value does not exist, cyan means the value is between 1 and 2 (no including 2) and indicating the country tends to be an importer, and magenta represents a relatively high value indicating an exporter. Three variables, i.e., the number of cyan points (say $n_{cyan}$), that of magenta points (say $n_{magenta}$) and the mean of MFPFD from the source to the country represented by the row (say $\overline{l_{0,i}^{\alpha}}$), are extracted to characterize each row (or column). Rows and columns of the matrix displayed in the figure are sorted by the sum of $n_{cyan}$ and $n_{magenta}$ in descending order, then by $\overline{l_{0,i}^{\alpha}}$.

In Fig.\ref{matrix_country-product}, we obtain an approximately right angle trapezoidal shape for cyan and magenta points. We can found that magenta points are concentrated in the upper portion of the trapezoidal shape, while the lower part is mainly cyan dots. It can be interpreted that the exporters of a wide spectrum of products are active countries in the international trade, which are located in the upper part of the figure and can be regarded as competitive and success countries according to \cite{Tacchella2012}. In contrast, bottom rows are those not active in the international trading, which can only export a limited variety of products and tend to be underdeveloped countries \cite{Tacchella2012}. Economies, which export few types of goods but import a large variety of products, also can occur in the upper part of the figure. Examples include Saudi Arabia and United Arab Emirates. Besides, we also find some abnormal countries, which appear as cyan horizontal lines in the top-right of the figure. It can be interpreted that these countries import many kinds of goods, which the vast majority of countries do not need to import. It may indicate that these economies are at high risk. A representative example is Yugoslavia in the year 2000.

\subsection{Node centrality}
\subsubsection{Countries' node centralities in a certain layer of product}
We calculate country centrality based on flow distances in a certain layer of product. The centrality of a country node can be computed as the average of distances from the node to all the other countries as given in \cite{Liangzhu2014}. However, if one of those distances is infinite, the node centrality will become infinite. Thus, the above computation method is infeasible for a sparse distance matrix in which most of the elements are infinite. Therefore, we propose a new definition of node centrality, called harmonic centrality, as
\begin{equation}{f}_{i} = 
\frac{N-1}{\frac{1}{f_{i1}}+\frac{1}{f_{i2}}+\cdots+\frac{1}{f_{i,i-1}}+\frac{1}{f_{i,i+1}}+\cdots+\frac{1}{f_{i,N-1}}+\frac{1}{f_{i,N}}},
\end{equation}
where $f_{i,j}$ ($j=1,\cdots, N$ and $j\neq i$) is the SMFDs between nodes $i$ and $j$, and $N$ is the number of common nodes in the flow network. Two special nodes, i.e., source node 0 and sink node $N+1$, are excluded, because here we are concerned about the distances of the node to the other common nodes, and the distances to the source and the sink have been depicted in the metric of trophic level.

Thus, the harmonic node centrality metric can avoid the infinite distance problem and well measure centralities of nodes. A smaller ${f}_{i}$ implies a more central position the node occupied in the flow network, because it has a smaller average distance to the other nodes. In Table \ref{table_centrality}, we list the top 5 and the bottom 5 countries with their ${f}_{i}$ in the decreasing order of ${f}_{i}$ in the money flow network of trading live bovine animals. The top 5 countries are all important hub nodes from the viewpoint of topology. 
\begin{table}
\caption{List of countries sorted by ${f}_{i}$}
\label{table_centrality}
\centering
\subfloat[Top 5 countries]{
  \begin{tabular}{ccc}
  \hline\hline
  Rank & Country & ${f}_{i}$ \\
  \hline
  1 & Germany & 2.53  \\
  2 & Netherlands & 2.74\\
  3 & Hungary & 3.05 \\
  4 & Australia & 3.10 \\
  5 & Italy & 3.15 \\
  \hline\hline
\end{tabular} }
\qquad
\subfloat[Bottom 5 countries]{
  \begin{tabular}{ccc}
  \hline\hline
  Rank & Country & ${f}_{i}$ \\
  \hline
  82 & Ecuador & 86.00\\
  83 & Mozambique & 86.00\\
  84 & Bahrain & 86.00\\
  85 & Qatar & 86.00 \\
  86 & Singapore & 86.00\\
  \hline\hline
\end{tabular} }
\end{table}




\subsubsection{Countries' node centralities for different products}
We compute and compare node centralities of countries in different layers of products. We establish a countries-products matrix containing the corresponding harmonic centralities. For example, the row for USA is $[3.95, 15.50, 5.03 , \cdots, 7.50]$, and the column for the commodity of live bovine animals is [86.00, n.a., n.a., 7.13, $\cdots$, n.a.], where n.a. means that the corresponding country do not appear in the trading of the commodity. We then transform the matrix into the countries-products matrix containing the rankings of countries for each type of commodity. 
Thus, the row of USA becomes $[15, 27, 11, \cdots, 4]$, which means that USA is ranked 15th, 27th, $\cdots$, and 4th respectively in each column, and the column for the product of live bovine animals turns into [79, n.a., n.a., 40, $\cdots$, n.a.].

\begin{figure}
\centering
\includegraphics[angle=0,width=5.7in]{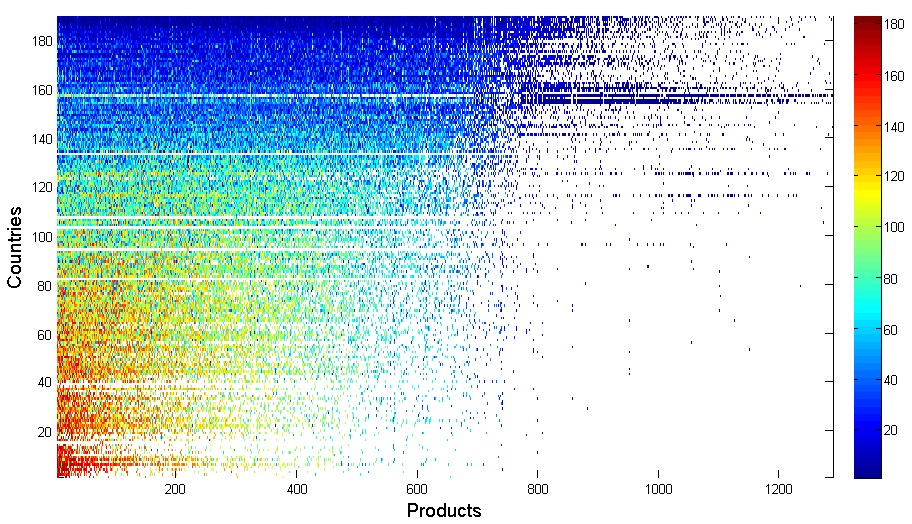}
\caption{Countries-products matrix reporting countries' harmonic centrality rankings in columns. Countries are sorted in ascending order by the average ranking, and products are arranged in descending order of the number of trading countries.}
\label{centrality_ranking_matrix}
\end{figure}

We illustrate the ranking matrix in Fig.\ref{centrality_ranking_matrix}, where rows are sorted by the mean of rankings, and columns are rearranged in descending order of the number of countries participating in the ranking. From the figure, we get an approximately right angle trapezoidal shape for colored dots, which is quite similar to the shape in Fig.\ref{matrix_country-product}. It can be interpreted that active countries which trade a wide spectrum of products are likely to have high average harmonic centrality; on the contrary, those having few types of trading products tend to have low average rankings. Most interestingly, in the figure there are several horizontal color bands, which are blue, cyan, yellow and red successively from top to bottom. This phenomenon can be named as \emph{centrality stratification}. It indicates that competitive countries tend to be in the center position in the trading of a large variety of products, while underdeveloped countries likely rank low in their limited varieties of trading products. We list the top 8 and bottom 8 economies with their corresponding average harmonic centrality rankings (say $\bar{f}_{i}$) and numbers of types of trading products (say $n$). It implies that $\bar{f}_{i}$ 
may be a good alternative indicator for countries' competitiveness from the perspective of node centrality, and countries need to promote their harmonic centrality rankings in the trading of various products for competitiveness enhancement. 

\begin{table}
\caption{List of economies sorted by the average harmonic centrality ranking. $n$ is the number of types of trading products and $\bar{f}_{i}$ the country's average harmonic centrality ranking.}
\label{table_centrality_ranking}
\centering
\subfloat[Top 8]{
  \begin{tabular}{cccc}
  \hline\hline
  Rank & Economy & $n$ & $\bar{f}_{i}$ \\
  \hline
  1 & Germany & 928  & 4.95 \\
  2 & USA & 992  & 5.47\\
  3 & France, Monac & 946  & 6.86\\
  4 & UK & 938 & 7.06\\
  5 & Italy & 917 & 8.04\\
  6 & China & 914 & 11.43\\
  7 & Spain & 912 & 12.13\\
  8 & Netherlands & 927 & 13.01\\
  \hline\hline
\end{tabular} }
\qquad
\subfloat[Bottom 8]{
  \begin{tabular}{cccc}
  \hline\hline
  Rank & Economy & $n$ & $\bar{f}_{i}$ \\
  \hline
  183 & Guinea-Bissau & 268& 126.72\\
  184 & Chad & 129& 126.94\\
  185 & Greenland& 299& 127.99\\
  186 & Burundi & 108& 128.60\\
  187 & Samoa & 123& 133.15\\
  188 & Occ. Pal. Terr & 61& 134.02\\
  189 & St.Pierre Mq & 68& 134.88 \\
  190 & CACM NES & 21& 137.86\\
  \hline\hline
\end{tabular} }
\end{table}





\subsection{Mean first-passage flow distances from the source to the sink}
\begin{figure}
\centering
    \subfloat[The distribution]
    {
       \begin{minipage}[t]{0.5\linewidth}
       \centering
       \includegraphics[angle=0,width=3.2in]{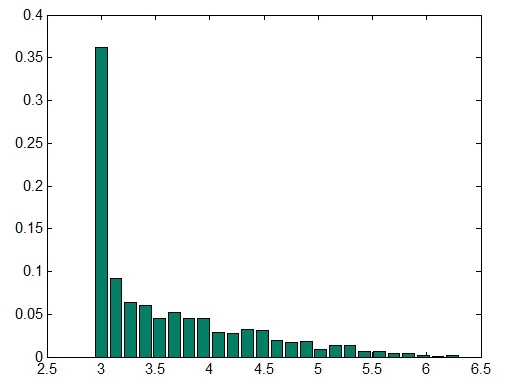}
       \vspace{0pt}
       \end{minipage}
       \label{SourceSink_distribution}
    }
    \subfloat[The cumulative distribution]{
       \begin{minipage}[t]{0.5\linewidth}
       \centering
       \includegraphics[angle=0,width=3.1in]{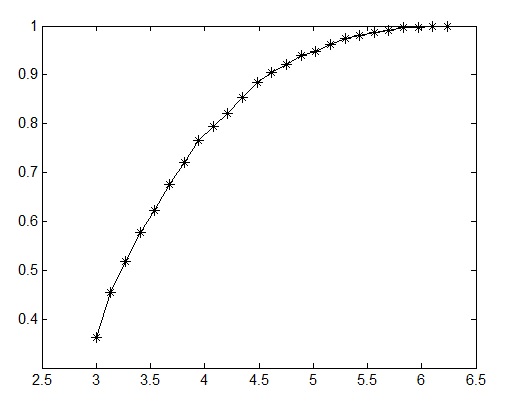}
       \vspace{0pt}
       \end{minipage}
       \label{SourceSink_cdf}
    }
    \caption{The distribution and cumulative distribution of MFPFDs from the source to the sink}
    \label{SourceSink_distribution_cdf}
\end{figure}

\begin{figure}
\centering
\includegraphics[angle=0,width=4.5in]{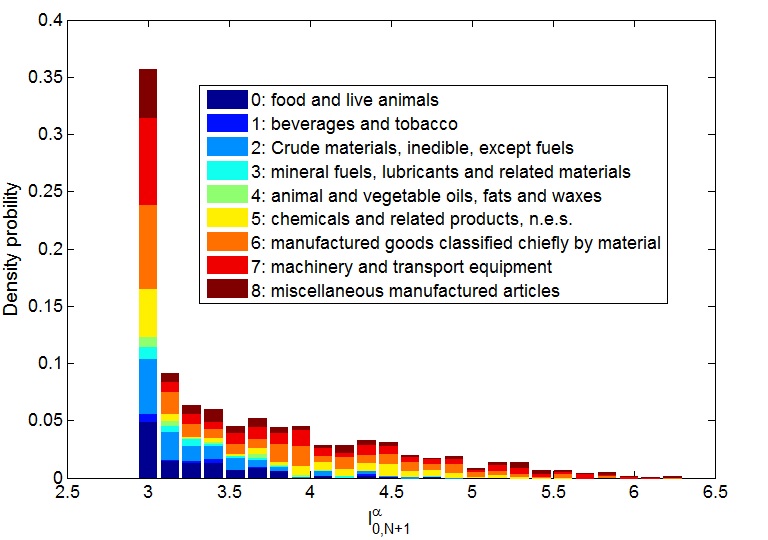}
\caption{The distribution of MFPFDs from the source to the sink for different categories of products}
\label{SourceSink_stacked}
\end{figure}

We calculate the mean first-passage flow distances (MFPFD) from the source to the sink (say $l_{0,N+1}^{\alpha}$) for money flow network in each layer. It indicates the average step that a random particle may jump from the source to the sink, and can be regarded as the flow length of the network in the layer. In Fig. \ref{SourceSink_distribution_cdf}, the distribution and cumulative distribution of $l_{0,N+1}^{\alpha}$ for all 1288 layers are shown. We find that over 27\% of $l_{0,N+1}^{\alpha}$ is 3, and the density of $l_{0,N+1}^{\alpha}$ decreases with the growth of the value of $l_{0,N+1}^{\alpha}$. The maximal one is 6.24. We further compare $l_{0,N+1}^{\alpha}$ among groups of layers containing different categories of products. According to the classification of products given by SITC4 \cite{SITC4}, the result is shown in Table \ref{table_MFPFD_groups}, where the category of 9 (commodities and transactions not classified) is ignored. From the table, we find that a group formed by categories 5 (chemicals and related products, n.e.s.), 7 (machinery and transport equipment), 6 (manufactured goods classified chiefly by material) and 8 (miscellaneous manufactured articles) have a significant higher average $l_{0,N+1}^{\alpha}$ than the group of categories 0 (food and live animals), 1 (beverages and tobacco), 2 (crude materials, inedible, except fuels), 3 (mineral fuels, lubricants and related materials) and 4 (animal and vegetable oils, fats and waxes). For simplicity, it can be concluded as manufactured products (categories 5, 6, 7 and 8) have a significant larger $l_{0,N+1}^{\alpha}$ than primary products (categories 0, 1, 2, 3, 4). This phenomenon also has been confirmed by the stacked distribution chart of $l_{0,N+1}^{\alpha}$ for different categories of products (Fig. \ref{SourceSink_stacked}), where blue bars are stacked on the left side by contrast with most yellow and red bars located on the right.



\begin{table}
\caption{MFPFDs from the source to the sink for different categories of products}
\label{table_MFPFD_groups}
\centering
  \begin{tabular}{clcccc}
  \hline\hline
  Code & Classification  & Average $l_{0,N+1}^{\alpha}$ & Minimal $l_{0,N+1}^{\alpha}$ & Maximal $l_{0,N+1}^{\alpha}$ \\
  \hline
  0 & Food and live animals & 3.35  & 3.00 & 5.86 \\
  1 & Beverages and tobacco  & 3.31  & 3.00 & 4.34\\
  2 & Crude materials, inedible, except fuels & 3.30  & 3.00 & 4.89 \\
  3 & \makecell[l]{Mineral fuels, lubricants and related\\ materials} & 3.28  & 3.00 & 4.80 \\
  4 & \makecell[l]{Animal and vegetable oils, fats and\\ waxes} & 3.34  & 3.00 & 4.07 \\
  5 & Chemicals and related products, n.e.s. & 3.78  & 3.00 & 5.61\\
  6 & \makecell[l]{Manufactured goods classified chiefly\\ by material}  & 3.71  & 3.00 & 6.17 \\
  7 & Machinery and transport equipment  & 3.75  & 3.00 & 6.12 \\
  8 & Miscellaneous manufactured articles  & 3.63  & 3.00 & 6.24 \\
  - & All products& 3.59  & 3.00 & 6.24 \\
  \hline\hline
\end{tabular}
\end{table}

\section{Conclusions and discussions}
In this paper, we use the approach of flow distances to explore the international trade flow system from the perspective of multi-layer flow network. The model of multi-layer flow network can well model the open flow system with multiple types of flows, where each layer contains one type of flows. We introduce the formal description of multi-layer flow networks (MFNs), and flow distances on the MFNs (e.g., the mean first-passage flow distance and the mean total flow distance), where a new flow distance called symmetric minimum flow distance is proposed. Then, we build the multi-layer flow network of international trade from two coupled viewpoints, i.e., the viewpoint of commodity flow and that of money flow.

Thus, based on the approach of flow distances, some interesting knowledge are discovered from the multi-layer flow network of international trade.

Firstly, countries' trading ``trophic levels'' is used to depict the positions that countries occupied in the international supply chain. From the distribution of trading ``trophic levels'', we find that countries can be divided into three groups: countries whose trophic levels are slightly bigger than or equal to 1, between 2 and 3, and larger than 3. Besides, since the viewpoints of commodity flow and money flow are coupled, some regularities can be found. For example, money flow network and commodity flow network for the same commodity have the same mean first-passage flow distance from the source to the sink.

Secondly, by comparing countries' trophic levels in different layers with different products, we find that exporters of a wide spectrum of products are active and competitive countries in the international trade, while countries export a limited types of commodities are inactive and tend to be underdeveloped countries. Besides, we find some countries import many kinds of goods, which the vast majority of countries do not need to import. This phenomenon may indicate that these economies are at high economic risk.

Thirdly, we propose a new node centrality called harmonic centrality for solving the problem of infinite distance. A smaller harmonic centrality indicates a more central position the node occupied. Then, we compare harmonic centralities of countries in different layers of products. It is interesting to find the phenomenon of centrality stratification. It means that competitive countries tend to be in the center position in the trading of a large variety of products, while underdeveloped countries likely rank low in their limited varieties of trading products.

Fourthly, we compute the mean first-passage flow distances from the source to the sink for different types of commodities, which can be regarded as the flow length of the network in each layer. We find that manufactured products have significant larger mean first-passage flow distances from the source to the sink than primary products.

Our findings demonstrate the effectiveness of the proposed model of multi-layer flow networks and the approach of flow distances.

\section*{Acknowledgements}
This work is supported by National Nature Science Foundation of China (No. 71271191, 71471165), the National Science \& Technology Pillar Program during the 12th Five-year Plan Period of China (No. 2012BAF12B11), Zhejiang Provincial Natural Science Foundation of China (No. Y15F020115) and Scientific Research Foundation for the Returned Overseas Chinese Scholars (Ministry of Human Resources and Social Security of China, 2013).
\section*{Additional information}
\textbf{Competing financial interests:} The author declares no competing financial interests.

\newpage

\end{document}